\documentclass[showpacs,twocolumn,amsmath,amssymb,pre,aps,floatfix,superscriptaddress]{revtex4}
\usepackage{epsfig}
\usepackage{graphicx,color}
\usepackage{amsmath}
\usepackage{dcolumn}
\newcommand{\changes}[1]{{\color{black}#1}}
\newcommand{\changesaa}[1]{{\color{black}#1}}
\newcommand{\changesbb}[1]{{\color{black}#1}}

\begin{document}
\title{Nano-scale mechanical probing of supported lipid bilayers with 
atomic force microscopy}
\author{Chinmay Das}
\email{c.das@leeds.ac.uk}
\affiliation{School of Physics and Astronomy, University of Leeds,
Leeds LS2 9JT, United Kingdom}
\affiliation{Unilever R $\&$D, Port Sunlight, Wirral, CH63 3JW, United Kingdom}
\author{Khizar H. Sheikh}
\affiliation{School of Physics and Astronomy, University of Leeds,
Leeds LS2 9JT, United Kingdom}
\affiliation{UCD Conway institute for Biomolecular and
Biomedical Research, Dublin, Ireland}
\author{Peter D. Olmsted}
\email{p.d.olmsted@leeds.ac.uk}
\affiliation{School of Physics and Astronomy, University of Leeds,
Leeds LS2 9JT, United Kingdom}
\author{Simon D. Connell}
\affiliation{School of Physics and Astronomy, University of Leeds,
Leeds LS2 9JT, United Kingdom}
\date{\today}
\begin{abstract}
We present theory and experiments for the force-distance curve $F(z_0)$
of an atomic force microscope (AFM) tip (radius $R$) indenting a supported
fluid bilayer (thickness $2d$).  For realistic conditions the force is
dominated by the area compressibility modulus $\kappa_A$ of the bilayer,
and,  to an excellent approximation, given by $F= \pi \kappa_A R
z_0^2/(2d-z_0)^2$.  The experimental AFM force curves from coexisting
liquid ordered and liquid disordered domains in 3-component lipid bilayers
are well-described by our model, and provides $\kappa_A$ in
agreement with literature values. The liquid ordered phase has a yield-like
response that we model as due to the breaking of hydrogen bonds.

\end{abstract}
\maketitle
\section{Introduction}
Atomic Force Microscopy (AFM) \cite{binnig.AFM.86} has become a
standard tool for imaging surfaces at high resolution and probing
local mechanical properties \cite{butt.AFM.review.05}. Force-distance
curves for indentation of AFM tips have been used to characterize the
mechanical properties of biological membranes
\cite{connell2006atomic,voitchovsky.afm.purplemem.06, costa.afm.2006}, and the usual
approach is to approximate the bilayer as an elastic solid undergoing
a Hertzian contact \cite{sneddon.afm.65, chadwick.afm.forcecurve.02,
  cheng.afm.06}. However, at physiological conditions most biological
membranes are in a fluid bilayer phase \cite{lipowski.mem.rev.91},
whose free energy is described by a bending modulus $\kappa$ and the
area compressibility modulus $\kappa_A$. These are experimentally
accessible through, for example, micropipette aspiration experiments
\cite{evans.micropipet.90}, which give the average value of the
elastic moduli over the whole vesicle. However, biological membranes
often have different local compositions, and thus different local
mechanical properties and physiological functions.

Despite the growing use of AFM to study lipid bilayers, the
flexibility of using it to measure local mechanical properties has not
been fully exploited. An AFM tip can bend a freely suspended membrane,
and compress a supported membrane. In recent work, Steltenkamp {\em et
  al.\ } \cite{steltenkamp06} showed how to extract the bending
modulus of lipid bilayers from AFM force-distance curves for bilayers
deposited over well defined sized holes (indentation of `nanodrums'),
in which they could safely ignore area compression due to the lack of
a supported surface.  Another issue neglected in previous AFM studies
is the double leaflet form of lipid bilayers, which is known to
influence the dynamics of fluctuations
\cite{seifert.memfluc.93}. Since an AFM tip induces an asymmetric
response in a supported bilayer \cite{XingSM2009}, the distinction
between the two leaflets will be important to accurately model the
mechanical response.

In this paper we consider the force-distance curves
\changesbb{obtained by indenting an AFM tip into a fluid bilayer
  supported on a solid substrate}.  The force-distance curves are
calculated from a static analysis of the deformation of the two
leaflets and differs from usual Hertzian result of the deformation of
elastic bodies.  We analyze experiments on a
dioleoyl-phosphatidylcholine (DOPC) - egg sphingomyelin (SM) -
cholesterol (CHOL) phase separated supported bilayer, which is a model
mixture representative of typical {\em in~vivo} membranes
\cite{veatch.ternary.05}.  \changesbb{ For certain composition ratios
  of the components, this system spontaneously phase separates into 
  coexisting liquid ordered (L$_o$, rich in SM and relatively thick
  because of strong nematic order in the acyl tails) and liquid
  disordered (L$_d$, rich is DOPC and relatively thin because of the
  more disordered tails) phases. } We show how to determine the area
compressibility moduli of the coexisting L$_o$ and L$_d$ phases of a
single sample, and find values in agreement with literature values.
To the best of our knowledge, this is the first time that the area
compressibilities of the two coexisting compositions in fluid bilayers
have been extracted directly. This technique should prove invaluable
for studying the composition dependence of mechanical properties in
lipid bilayers, and can be easily extended to consider more complex
interactions between AFM tip and the bilayer.

\section{Theory} 
\begin{figure}[htbp]
\centerline{\includegraphics[width=7.5cm, clip=true]{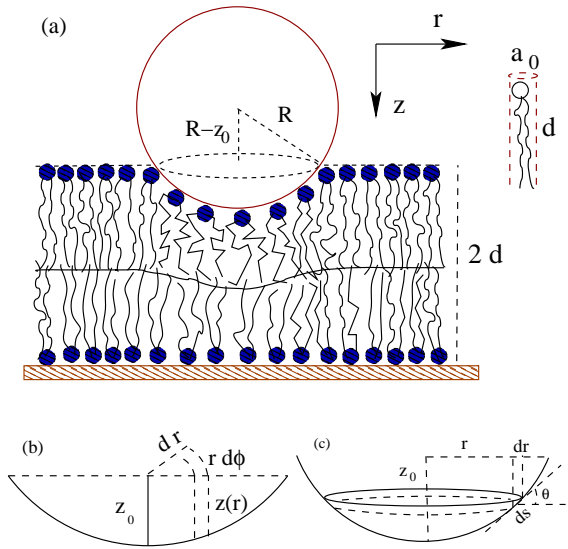}}
\caption{(color online) (a) Schematic geometry of an AFM tip of radius 
  $R$ indenting
  a fluid bilayer of thickness $2 d$ by an amount $z_0$. The leaflet
  dividing surface at $h_b(r)$ is shown as a solid line. (b)
  Cylindrical volume elements of depth $z(r)$ at the distance $r$ from
  the center of the tip. (c) The area $2\pi r\,ds$ of such an element
  in contact with the lipid. }
\label{fig.afm.sch}
\end{figure}

We consider a supported fluid lipid bilayer of
thickness $2 d$ probed by an AFM tip of radius $R$ in contact mode,
which measures the force as a function of the depth $z_0$ from the
unperturbed surface of the layer (see Fig.~\ref{fig.afm.sch}(a)). At
typical AFM speeds the viscous forces are negligible.  We begin by
assuming a hard contact interaction between the tip and the membrane.
Electrostatic repulsion from the charged double layers and van der
Waals attraction are included later in the paper when comparisons are
made with experiments.  Since the lipid bilayer is not anchored it
remains tension free.  We assume that the volume is conserved at the
molecular level: as the tip penetrates the bilayer it occupies a
volume $\delta V$, so that $\delta N=\delta V/(a_0 d)$ lipids are
expelled into the surrounding bilayer. Here, $a_0$ is the area per
lipid in the absence of the AFM tip. The surface area increases by
$\delta A$, due to the curved spherical surface of the AFM tip, and
the increase in area per head group $\delta a \equiv a - a_0$ is given
by
\begin{equation}
\delta a  = \frac{A + \delta A}{N - \delta N} - a_0
= a_0 d \left( \frac{A + \delta A}{V - \delta V} \right) - a_0.
\end{equation}
This increased area induces an elastic cost due to the stretching
elasticity of the lipid leaflets. We calculate this not by averaging
over the entire spherical cap, but by considering small increases in
radius $dr$, and evaluating $\delta a/a_0$ at each $r$
(Fig.~\ref{fig.afm.sch}). 

We assume that both leaflets have the same area per lipid $a_0$ and
stretching modulus $\kappa_A/2$ for lipid head groups on a flat
surface.  This should be valid in the absence of specific interaction of
the lipid with the substrate, \changes{although experiments have shown
  that the surface often does have specific interactions
  \cite{XingSM2009}.}  The head groups in the top leaflet are forced
to lie on a curved surface below the AFM tip. This affects both the
area/lipid, and the stretching modulus for the lipids on the top
leaflet.  We model the local lipid free energy as a sum of a surface
energy and a harmonic tail stretching, $g(L) \simeq \gamma a_L/\cos
\theta + \alpha/ a_L^2$, where $\gamma$ is a surface tension, $\alpha$
penalizes tail stretching, and $a_L$ is the projected area for leaflet
thickness $L$.  Here, $\theta (r) = \sin^{-1} (r/R)$
(Fig.~\ref{fig.afm.sch}c) is the tilt angle of the lipid surface.
Minimizing $g$ at fixed lipid volume $v \simeq L a_L$ leads to an
effective stretching modulus $\tilde{\kappa}_A/2 \simeq \kappa_A/2\,
\sec^{2/3}\theta$ and an effective area per lipid $a_0^t=a_0\,
\cos^{1/3}$ for the top leaflet.

Using the modifications due to the curved surface for the head-groups in 
the top leaflet, the excess free energy due to the increase
in the area per lipid during indentation is 
\begin{equation}
G(z_0) = \frac{\kappa_A}{4} \int_S  d^2r\left[
  \sec^{2/3} \theta(r) \left( \frac{\delta a}{a_0} \right)_{t}^2 +
 \left( \frac{\delta a}{a_0} \right)_{b}^2 \right],
\label{eq.fen.entropic}
\end{equation}
where $t$ and $b$ refers to the top and bottom leaflets and the
integration extends over both leaflets. The lower leaflet will
generally deform to accommodate the large energy change due to
removing too many lipids from the upper leaflet. We let the lower and
upper leaflets have thicknesses $h_b(r)$ and $h_t(r)$ respectively,
with $h_b(r)+h_t(r)=2 d - z_0 + R[1-\cos\theta(r)]$.  The area changes
at each radius $r$ are given by (Fig.~\ref{fig.afm.sch})
\begin{subequations}
\label{eq:1}
\begin{eqnarray}
 \left( \frac{\delta a}{a_0^t} \right)_{t}^2  &=& 
  \left[\frac{d\sec^{1/3} \theta(r)}{h_t(r)} 
- 1\right]^2,    \\ 
 \left( \frac{\delta a}{a_0} \right)_{b}^2  &=& 
\left( \frac{d}{h_b(r)} - 1 \right)^2.
\end{eqnarray}
\end{subequations}
The measure is $d^2r=r\,d\phi dr$.
The dividing surface $h_b(r) $ is determined by minimizing the
free energy  at each $r$. 
\changes{For equal stretching moduli in both leaflets, an explicit
  solution for $h_b(r)$ is possible for small tilt angle,}
\begin{equation}
h_b(r) = \frac{2 d -z_0 + R \left(1 - \cos \theta(r) \right)}
{1 + \sec^{1/3} \theta}.
\label{eq.zm.approx}
\end{equation}
\changes{For realistic values $z_0\sim2$nm, $d\sim3$nm and
  $R\sim10$nm}, this approximation introduces less than $0.1\%$ error
in $h_b(r)$ for the entire range of $r$. We use this approximation in the
rest of the paper to derive analytic expressions for the free energy
and force.

Using $h_m(r)$ from Eq.~\ref{eq.zm.approx}, the free energy is
\begin{align}
\frac{2 G(z_0)}{\pi \kappa_A R^2} &= \int_{1-\frac{z_0}{R}}^1
x \left(1 + x^{-2/3} \right) \times \nonumber \\ 
& \left[ \frac{ \left(1 + x^{1/3} \right)}
{x^{1/3} \left[2 - \frac{z_0}{d} + \frac{R}{d} (1 - x) \right]}
- 1 \right]^2 dx,
\label{eq.fen}
\end{align}
and the force  on the AFM tip is given by 
$F = {\partial G}/{\partial z_0}$.  \changesbb{We use the numerical force derived from Eq.~\ref{eq.fen.entropic} when performing fits to the data}. For small penetrations $z_0$ the force can be written as
\begin{equation}
F\simeq \frac{\pi \kappa_A R}{4}\left[1 + \frac{d}{3R} +
  \left(\frac{d}{3R}\right)^2\right] \left(\frac{z_0}{d}\right)^2 + {\cal O}(z_0^3)\ldots
\label{eq.frc.lead}
\end{equation}
\changes{A surprisingly simple function that fits the entire
  experimental range of forces, correct to within a few percent for
  $R=3d$ and much better for larger $R$, is
\begin{equation}
F= \frac{\pi \kappa_A R}{4}\left(\frac{2z_0}{2d-z_0}\right)^2.
\label{eq.frc.simple}
\end{equation} }
\changes{The force diverges as $z_0$ approaches $2 d$ because the
area/lipid diverges in order to preserve molecular volume. The
quadratic free energy (Eq. \ref{eq.fen.entropic}) is no longer valid
there.  Experimentally this divergence is preempted by pore
formation (see below).}

\begin{figure}[htbp]
\centerline{\includegraphics[width=7cm, clip=true]{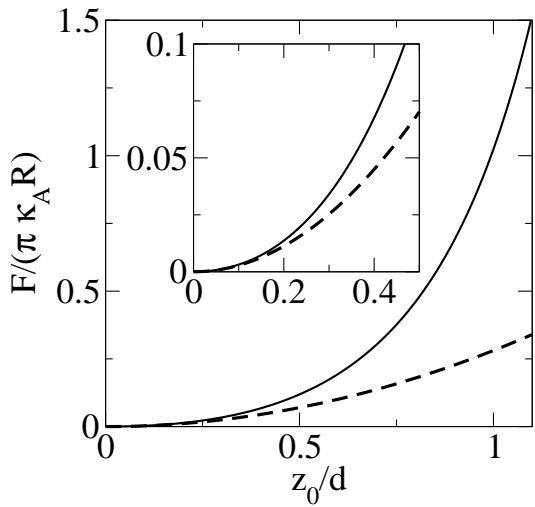}}
\caption{ Scaled force $F/(\pi \kappa_A R)$ as a function of tip depth
  $z_0/d$ (solid line) for tip radius $R/d = 3$, according to
  Eq.~\ref{eq.fen}. The dashed line shows only the leading quadratic
  term in $z_0/d$, Eq.~\ref{eq.frc.lead}. Inset: Behaviour at small
  $z_0/d$. \changesbb{On this scale the approximation of
  Eq.~(\ref{eq.frc.simple}) is indistinguishable from the numerical
  solution of Eq.~(\ref{eq.fen}).}}
\label{fig.fenfrc.th}
\end{figure}

For comparison, the contact force between two solid (elastic) bodies
much larger than the radius of contact (Hertzian contact) is $F \sim
z_0^{3/2}$ \cite{landau.elasticity}.  More relevant for AFM
experiments, the force to indent a finite elastic layer scales as
$F\sim R^2 z_0^3 / d^3$ if the layer is bonded to the substrate and
$F\sim R z_0^2 /d$ if the layer can slip
\cite{chadwick.afm.forcecurve.02}.  The response of fluid bilayers in
eqn.~\ref{eq.frc.lead} scales differently than all of these scenarios,
and force is proportional to the area compressibility modulus instead
of the Young's modulus.  For realistic experimental values the region
of validity for this quadratic behavior is limited, as shown in
Fig.~\ref{fig.fenfrc.th}(b). 

\section{Experiments} 
\begin{figure}
\centerline{\includegraphics[width=7.5cm, clip=true]{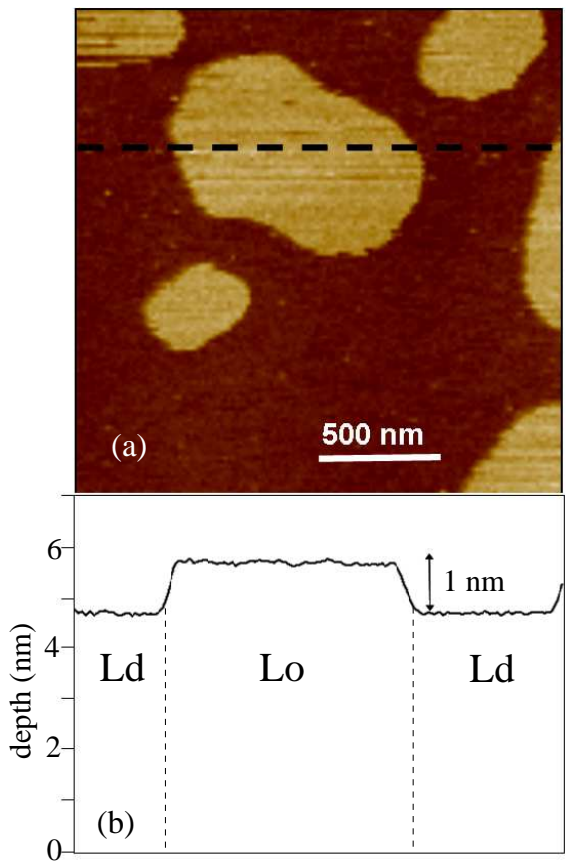} }
\caption{(color online) Phase separated lipid bilayer with liquid ordered and
  liquid disordered domains. (a) Tapping mode AFM image showing the
  height profile of the bilayer. (b) One dimensional section along the
  dashed line in (a).}
\label{fig.twophase}
\end{figure}

To test the theory we performed experiments on a supported bilayer
comprising DOPC, SM and CHOL \changesbb{at overall molar ratios
  DOPC:SM:CHOL=40:40:20}.  At room temperature this system phase
separates into coexisting \changesaa{DOPC-rich} liquid disordered and
\changesaa{SM-rich} liquid ordered domains (Fig.~\ref{fig.twophase}).
\changesbb{The hydrocarbon tails have large nematic order in the L$_o$
  phase, leading to a thicker bilayer and higher area compressibility
  modulus. In contrast, the tails have lower nematic order in the
  L$_d$ phase with concomitant smaller thickness and lower moduli.}
DOPC, CHOL (purchased from Sigma) and egg SM (purchased from Avanti)
were dissolved in chloroform, dried under a stream of argon for 30
minutes, and then vacuum desiccated for 30 minutes. The lipid was
resuspended in PBS buffer at pH 7.4 to a concentration of 1 mg/ml by
vortexing.  To make small single unilamellar vesicles (SUVs), the
cloudy lipid suspension was tip sonicated (IKA, U50) at less than
$5^{\circ}\,\textrm{C}$ for 25 mins (until the solution became clear).
The mica (Agar Scientific Ltd.)  surface was incubated with the SUVs
at \changesaa{at 50$^{\circ}$~C and cooled down to room temperature in
  a incubator over 15 minutes.  After 1h, the sample was} gently
rinsed with PBS buffer to remove any excess vesicles.

\begin{figure}[htbp]
\centerline{\includegraphics[width=7.5cm, clip=true]{./dopcdata.eps} }
\hbox{}\vskip-7cm\hskip3.5cm\includegraphics[height=2.8cm,clip=true]{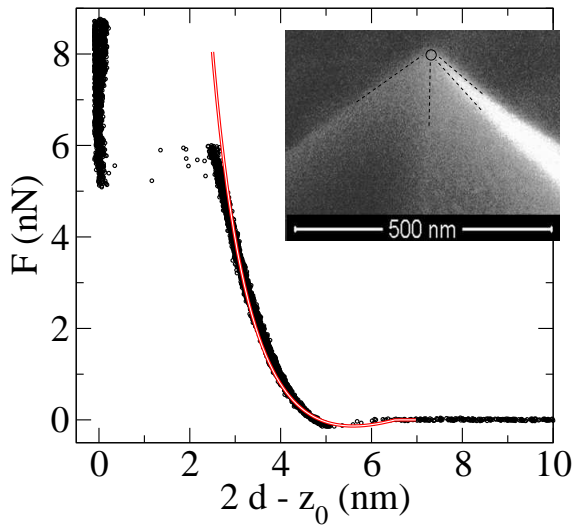}
\hbox{}\vskip3.5cm
\caption{Force-distance curve for a DOPC rich bilayer in the liquid disordered phase.  The data
  points are from AFM experiments and the line is a fit for the
  theoretical prediction with $\kappa_A = 0.12\,\textrm{N/m}$.
  Inset: SEM image of the tip to measure the tip radius. The dashed
  lines are along the pyramidal face edges. The circle drawn at the
  end of the tip has a radius $10\,\textrm{nm}$.  }
\label{fig.ex.data}
\end{figure}

\changesbb{Force measurements were performed at
  $27^{\circ}\;\textrm{C}$ in PBS buffer using a Nanoscope IV
  Mulitmode AFM (Veeco) equipped with a temperature control stage,
  using cantilevers (NP, Veeco) with nominal spring constants of 0.12
  N/m. Spring constants were measured using the thermal noise method
  \cite{hutter.afmex.93} in air, and optical lever sensitivity
  determined against a clean mica surface.The force curves analysed in
  this paper were all taken from a single force-volume map of the
  phase separated bilayer shown in Figure.~\ref{fig.ex.data}, and
  exported using Nanscope software v5.12r30.}

Scanning electron microscopy (Camscan series III, FEG-SEM operating at
5~kV with magnification 160k) was used to measure the tip
radius. Inset of Fig.~\ref{fig.ex.data} shows the tip image with
\changes{dashed lines along the edges of the four pyramidal
  faces}. The end of the tip can be approximated as spherical. The
drawn circle (Fig.~\ref{fig.ex.data} inset) has a radius of 10~nm. The
contrast of the image is poor. Consequently the uncertainty of the
exact value of the radius is large. In our analysis we consider the
tip radius to be $R=10 \pm 5\,\textrm{nm}$.

Fig.~\ref{fig.twophase} shows a tapping mode image of the bilayer
along with a one dimensional cross section. There is a $\sim
5\,\textrm{nm}$ thick L$_d$ matrix enclosing $\sim 6\,\textrm{nm}$
thick L$_o$ domains \changesaa{(The heights reported here include the
  thickness of any water layer between the bilayer and the mica
  surface)}.  \changesbb{The composition of the two phases were
  determined by following the tie lines on the ternary phase diagram,
  which were calculated using Atomic Force Microscopy
  \cite{simon.ternary.08}:
\begin{subequations}
\begin{align}
L_d:&&(\textrm{DOPC:SM:CHOL})&=(68:27:5)\label{eq:phi.Ld}\\
L_o:&&(\textrm{DOPC:SM:CHOL})&=(3:71:26)\label{eq:phi.lo}
\end{align}
\end{subequations}
} \changesaa{Phase diagrams on similar ternary mixtures have been
  calculated using NMR, and the compositions of the liquid-disordered
  and liquid ordered phases are similar \cite{veatch06}.  The
  uncertainty in the compositions from placement of the tie lines is
  estimated to be less than 2\% of the quoted values. } For both L$_o$
and L$_d$ phases, force curves from contact mode AFM were used from at
least 10 different measurements from different points within different
regions (`patches', as in Fig.~\ref{fig.twophase}) of the sample.

Fig.~\ref{fig.ex.data} (symbols) shows the force curve for the DOPC
rich  bilayer in the liquid
disordered phase.  Besides the stretching contribution considered so
far, the tip experiences an attractive force due to van~der Waals
interaction and a short range repulsive force due to \changesbb{the electric double layers}
on the tip surface and the membrane top surface.  In principle the
van~der Waals interaction can be calculated from a knowledge of the
dielectric constants of the tip, membrane, and the PBS buffer
\cite{israelachvili.book}.  Similarly, the repulsive interaction can
be estimated by knowing the detailed charge distribution and solving
Poisson-Boltzmann equation. Phenomenologically, we model the van~der
Waals attraction as an interaction energy of the form $-A/\xi^6$
between volume elements of the tip and the membrane separated by a
distance $\xi$.

Since these forces are short-ranged, we consider the tip as a sphere
and for the volume integration over the membrane, consider the
membrane as infinitely thick.  We further assume that the repulsive
interaction is strong enough to avoid adsorption. As the tip
approaches the bilayer, the bilayer deforms.  The extent of the
deformation is governed by the minimum of the stretching free energy
and the long range interactions (van~der Waals and screened
Coulomb). We assume that the deformation can be modeled as hard
interaction from a tip with radius $R_c$ larger than the physical tip
radius $R$.
\changes{Although $A$ is poorly
  known and depends on the detailed dielectric properties of the
  membrane, its precise value only changes $R_c$ and controls the details
  of the force near contact. For deeper contact the force is
  overwhelmingly dominated by the stretching modulus $\kappa_A$, so that
the force-distance curves yield the same
  $\kappa_A$, independent of $A$. }

The drawn line in Fig.~\ref{fig.ex.data} shows the fit from our
theoretical analysis, using a downhill simplex method \cite{book.nr}
to minimize the mean square fractional deviation of the prediction
from the experimental data over the fitted range.  The best fit for
the compressibility modulus for $R=10\,\textrm{nm}$ is
$\kappa_A=0.12\,\textrm{N/m}$.  Because of the uncertainty in the
tip radius, the range of $\kappa_A$ for $R$ between $5\,\textrm{nm}$
and $10\,\textrm{nm}$ is between $0.25 \,\textrm{N/m}$ and
$0.08\,\textrm{N/m}$, respectively.  Our estimate compares well with the literature
values $\kappa_A=0.13-0.6\,\textrm{N/m}$ from osmotic pressure
measurements \cite{tristram-nagle.osmotic.dopc.98} and
$\kappa_A=0.18\pm0.04\,\textrm{N/m}$ from micropipette aspiration of
GUVs \cite{fa.guv.dopc.07} made of pure DOPC. Our model provides an
excellent fit until $2d - z_0 \simeq 2.5 \textrm{nm}$, at which point
the elastic energy of the deformed bilayer overcomes the cost of
forming a hole \cite{butt.afm.02} and the tip abruptly penetrates the
full bilayer. 

\begin{figure}
\centerline{\includegraphics[width=7.5cm, clip=true]{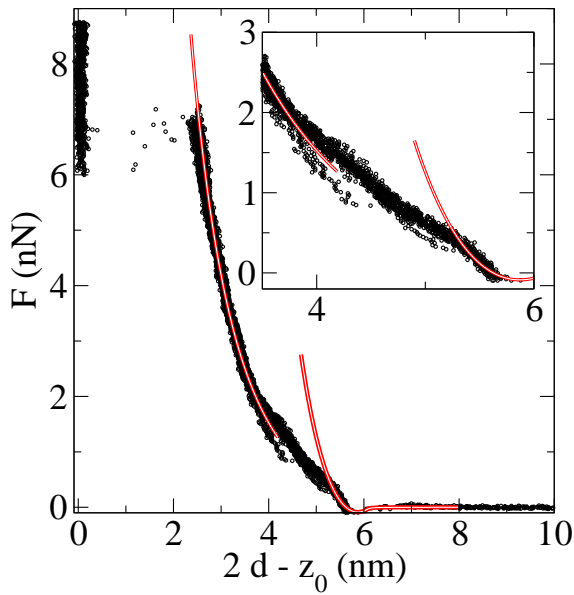} }
\caption{ (color online) Force-distance curve from AFM (symbols) in the SM rich
  liquid ordered phase superposed with two separate theoretical fits
  (lines, \changesbb{using Eq.~\ref{eq.fen}}) involving two different $\kappa_A$ at small and large tip
  penetrations $z_0$.  Inset: Closeup of the crossover region. }
\label{fig.ex.SM}
\end{figure}
\section{Response of Liquid Ordered Domains}
The AFM force curves for the SM rich liquid ordered phase are
qualitatively different from those in the coexisting liquid disorded
phase (Fig.~\ref{fig.ex.SM}). The initial deformation ($5\,\textrm{nm}
\le 2d - z_0 \le 6\,\textrm{nm}$) shows a high modulus
\changesbb{consistent with the tightly packed character of the L$_o$
  phase}.  Around $2d - z_0 \simeq 5\,\textrm{nm}$ the response shows
a crossover to a much lower modulus. The symbols in
Fig.~\ref{fig.ex.SM} are from 12 separate force-distance measurements.
While the experimental data fall on the same curves away from the
crossover region, the transition from stiff to soft behavior occurs at
different values of $z_0$, \changes{which may be due to either the
  stochastic behaviour of an activated event or flucuations in
  composition from region to region.} 

\changesbb{We first attempt to model these force curves as due to an
  effective stretching modulus that differs for small and large
  penetrations far from the crossover region. Hence we fit the data
  at small penetration ($2 d - z_0 > 5.3\,\textrm{nm}$) and large
  penetration ($2 d - z_0 < 3.5\,\textrm{nm}$), with effective
  stretching moduli according to Eq.~(\ref{eq.fen}).} For
$R=10\,\textrm{nm}$, the small $z_0$ fit gives $\kappa_A = 1.1
\,\textrm{N/m}$.  Recent experiments on a bovine brain SM and CHOL
equimolar mixture found $\kappa_A = 2.1\pm0.2\,\textrm{N/m}$
\cite{rawicz.sm.areacomp.08}.  Since egg SM (16:0 SM) has shorter
fatty acid chains than does bovine SM (18:0 SM), and the current
composition has comparatively smaller amounts of CHOL, we expect the
membrane to be softer (smaller $\kappa_A$), as found.

The large penetration ($z_0$) region has a stretching modulus
$\kappa_A =0.05\,\textrm{N/m}$, \changesbb{which is much closer to
  that of the L$_d$ phase shown in Fig.~\ref{fig.ex.data} than the
  unperturbed L$_o$ phase.  The AFM tip forces the bilayer immediately
  below it to decrease in thickness, which thus destroys the strong
  nematic order of the L$_o$ phase and induces a yielding or phase
  transition of the L$_o$ phase into an L$_d$ phase.  It is likely
  that the composition of this induced L$_d$ phase differs from that
  of the L$_d$ phase that characterizes equilibrium coexistence far
  from the AFM tip (Eq.~\ref{eq:phi.Ld}), because of slow kinetics of
  composition changes under the AFM tip. Our separate fits to extract
$\kappa_A$ suffer from narrow available fit window
($\sim0.3\,\textrm{nm}$) for small $z_0$ and the lack of small force
data for the large $z_0$ fit. Also, this procedure does not elucidate
the reason for two distinct elastic regions separated by a crossover. }

\begin{figure}
\centerline{\includegraphics[width=7.5cm, clip=true]{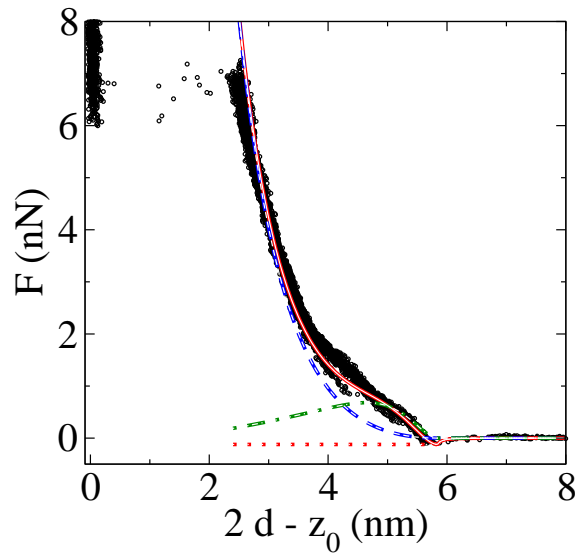} }
\caption{(color online) Force-distance curve from AFM (symbols) in the SM rich
  liquid ordered phase superposed with a \changesbb{microscopically-motivated  fit that accounts
  for a separate energetic contribution from hydrogen-bond breaking
  (solid line, based on Eq.~\ref{eq.fen.hbnd}).}  Also shown are the separate contributions from the
  van~der Waals interaction (dotted line), from the hydrogen bonds
  (dot-dashed line) and the area compressibility term (dashed line).}
\label{fig.ex.hbnd}
\end{figure}
\changes{To understand the qualitatively difference force responses of
  the L$_o$ and L$_d$ phases, we propose a
  \changesbb{microscopically-motivated} model.  SM has both hydrogen
  bond donor and acceptor groups and is known to form inter-SM
  hydrogen bonds \cite{mombelli.hbnd.03, rog.hbnd.06}.  The free
  energy in the L$_d$ phase, as represented in
  Eq.~\ref{eq.fen.entropic}, is dominated by solvent and tail packing
  entropies. Hence, to describe the L$_o$ phase we separately include
  the short range energy of hydrogen bond breaking through a simple
  Morse potential:} $U(b) = E_D \left[1 - \exp\left(-(b-b_0)/\lambda_m
  \right)\right]^2$, where $b$ is the separation between the donor and
acceptor group and $b_0$ is the equilibrium separation.  For typical
hydrogen bonds the dissociation energy $E_D \sim 2 - 7 \,
\textrm{KCal/mol}$ and the range $\lambda_m \sim 0.02 - 0.07 \,
\textrm{nm}$ \cite{gao.hbnd.84, thierry.hbnd.93}. For small changes in
area/lipid and affine deformation the contribution to the free energy
from distortion of hydrogen bonds is approximately
\begin{align}
  \label{eq.fen.hbnd}
G_{\textrm{\tiny{HB}}} (z_0) &= e_{\textrm{\tiny HB}}\!\! \int\!\! d^2 r\!\left\{
\frac{h_t(r)}{d} \left[1 - e^{-\frac{1}{\lambda}\left(\frac{\delta a}{a_0} \right)_t}
 \right]^2  \right. \nonumber \\
& + \left.
\frac{h_b(r)}{d} \left[1 - e^{-\frac{1}{\lambda}\left(\frac{\delta a}{a_0}\right)_b} \right]^2
\right\}.
\end{align}
Here, $e_{\textrm{\tiny HB}}$ is inter-lipid hydrogen bond
dissociation energy per area and $\lambda \sim 3^{1/4}
\lambda_m/\sqrt{a_0}$ for hexagonal arrangement of the lipids. As
before, the total free energy,  now comprising contributions from
Eqs.~(\ref{eq.fen}, \ref{eq.fen.hbnd}), is minimized at each $r$ to find
  the dividing surface between the leaflets and the force is
  calculated from $F=\partial G/\partial z_0$.

\changesbb{In the limit of small penetrations this model gives an effective stretching modulus $\kappa_{A}^{\textrm{eff}}$ in the L$_o$ phase of 
\begin{equation}
\kappa_{A}^{\textrm{eff}}=\kappa_A+4\frac{e_{\tiny HB}}{\lambda^2},
\end{equation}
where $\kappa_A$ is thus the stretching modulus of the L$_d$ phase
that is left after the L$_o$ phase has been destabilized and there is
no remaining hydrogen bond contribution.  The fit to the data is shown
in Fig.~\ref{fig.ex.hbnd}. The stochastic nature of the force curves
near the rupture point ($2d-z_0\simeq5.7$), limits the ability to
obtain excellent fits. Our fit gives $\kappa_A = 0.13\,\textrm{N/m}$,
$e_{\textrm{\tiny HB}}=0.006\,\textrm{N/m}$ and $\lambda=0.1$.}

Assuming an area per lipid $a_0\sim 0.6 \,\textrm{nm}^2$, \changes{the fitted
  value $\lambda$ implies that the range of the Morse potential is
  $\lambda_m \sim 0.06 \,\textrm{nm}$.}  Simulations show about 0.5
hydrogen bonds per lipid in SM bilayer \cite{mombelli.hbnd.03}.
Assuming an average hydrogen bond energy of
$3.5\,\textrm{KCal/mol}$, our value for $e_{\textrm{\tiny HB}}$ give 0.4 hydrogen bonds broken
per lipid.  The initial deformation is dominated by the contribution
from the hydrogen bonds, and the corresponding force curve leads to an
 area compressibility modulus ${\kappa}^{eff}_A \simeq 2.7\,\textrm{N/m}$.
 
\section{Discussion}  
 We have assumed a static force
 response, despite typical tip velocities $v_{tip}\simeq 10^2$nm/s.
 We can estimate the correction due to finite tip velocity by considering the
 dissipation from two dimensional viscosity $\eta$ of the lipid layer.
The dissipative force is found to be
\begin{equation}
F_D(z_0) = \frac{\eta \pi z_0 (2 R - z_0)}{2 d^2} v_{tip}.
\end{equation}
The two dimensional shear viscosity for fluid bilayers is expected to be of
the order of $10^{-10}$N-s/m \cite{sickert.viscosity.03}, leading to 
$F_{D}\simeq10^{-8}$ nN, much smaller than the
 elastic contributions. Hence our static approach is sufficient to describe the AFM
force-distance curves on fluid lipid layers.

In our calculations we have assumed that the two leaflets have the
same area compressibility and preferred area per head group. 
This may not always be the case, because of surface interactions
\cite{XingSM2009}; for example,  supported bilayers often have
different melting temperatures than their counterparts in giant unilamellar
vesicles. Incorporation of asymmetric membranes into the model is
straightforward, although more complex. We have also neglected splay
or bending energies.  Part of the elastic cost of this is already
included in the increased area/lipid against the curved surface, in
Eq.~\ref{eq.fen.entropic}, but there may also be an additional
negligible free energy cost due to the splay of the lipid tails,
through the bending modulus of each leaflet.

\changesaa{In our analysis, the initial deformation for the 
L$_{o}$ phase is described in terms of the stretching of 
hydrogen bonds. This can be explicitly tested by performing
experiments with varying concentrations of SM or using chemicals
that disrupt hydrogen bonds. However, this is beyond the scope of 
the present work.}

\changes{Evidently, local applied pressure can melt the liquid ordered
  phase into the thinner L$_d$ phase, which is not surprising.  We
  have proposed an explicit microscopic mechanism in terms of breaking
  hydrogen bonds that are implicated in stabilizing \changesbb{L$_o$} phase. An
  alternative and more general description could include a Landau
  theory for the free energy of the L$_o$-L$_d$ phase transition, with
  local pressure $p$ added as an external field to destabilize the L$_o$
  phase, $\Delta G \sim p \psi $, where $\psi$ is an order parameter
  proportional to thickness whose value decreases upon a transformation to  the L$_d$ phase
  \cite{komura04}. The phase transformation would then occur first at
  constant composition, and then one may expect the composition to
  change slowly as the external force changes the local preference for
  the different lipid species. The subject of kinetics and
  composition as a function of applied pressure is interesting and
  important, and we leave this for further work.}

In summary, we have presented, and validated by experiments, a theory
for describing the force distance $F(z_0)$ relationship for AFM
experiments on fluid bilayers, which leads to a remarkably simple
expression for $F(z_0)$, Eq.~(\ref{eq.frc.simple}).  This provides a
method for finding the area compressibility modulus and the amount of
inter-lipid hydrogen bonds of fluid bilayers. The agreement with the
existing literature values for the area compressibility is
excellent. \changes{The main uncertainty in our prediction is due to
  the uncertainty in the tip radius $R$. However, the simple linear
  dependence on $R$ means that relative measurements taken with the
  same tip can be compared very accurately.}

\acknowledgments The authors thank A. Ferrante, R. Marriot, M. Noro
and B. Stidder for useful discussions.  This work was supported by
Yorkshire Forward through the grant YFRID Award B/302. CD acknowledges
SoftComp EU Network of Excellence for financial support.

\bibliography{afm}

\end{document}